\documentclass[final,5p,twocolumn,authoryear]{elsarticle}

\usepackage{graphicx}
\usepackage{subfigure}
\usepackage{epsfig}
\usepackage{natbib}
\usepackage{amssymb}
\usepackage{amsthm}
\usepackage{lmodern}
\usepackage{geometry}
\usepackage {psfrag}
\usepackage{color}
\usepackage{lineno}
\usepackage{pifont}

\journal{Journal of Theoretical Biology}

\begin{document}

\begin{frontmatter}

\title{Are leaves optimally designed for self-support? An investigation on giant monocots.}

\author[label1,label2]{Lo\"\i c Tadrist \fnref{label4,label5}}
\author[label3]{Baptiste Darbois Texier\fnref{label5}}
\address[label1]{LadHyX, Department of Mechanics, \'Ecole Polytechnique-CNRS, 91128 Palaiseau, France}
\address[label2]{Microfluidics Lab, Department of Aerospace and Mechanics, University of Li\`ege, Li\`ege 4000, Belgium}
\address[label3]{GRASP, Physics department, University of Li\`ege, B4000 Li\`ege, Belgium}

\fntext[label4]{{loic.tadrist@ulg.ac.be}}
\fntext[label5]{{Both authors contributed equally to the work}}

\begin{abstract}
Leaves are the organs that intercept light and create photosynthesis. Efficient light interception is provided by leaves oriented orthogonal to most of the sun rays. Except in the polar regions, this means orthogonal to the direction of acceleration due to gravity, or simply horizontal.
The leaves of almost all terrestrial plants grow in a gravity field that tends to bend them downward and therefore may counteract light interception.
Plants thus allocate biomass for self-support in order to maintain their leaves horizontal. To compete with other species (inter-species competition), as well as other individuals within the same species (intra-species competition), self-support must be achieved with the least biomass produced. This study examines to what extent leaves are designed to self-support. We show here that a basic mechanical model provides the optimal dimensions of a leaf for light interception and self-support. These results are compared to measurements made on leaves of various giant monocot species,especially palm trees and banana trees. The comparison between experiments and model predictions shows that the longer palms are optimally designed for self-support whereas  shorter leaves are shaped predominantly by other parameters of selection.     
\end{abstract}


\end{frontmatter}


\begin{table}
\begin{center}
\begin{tabular}{ll}
\hline
Symbols&Parameters \\
\hline
$\theta$ & Maximal angle of deflection on the palm\\
$V$& Available volume of biomass\\
$S$& Leaf surface\\
$M$& Total mass of the leaf\\
$M_e$ & Extra mass attached to the leaf free-end\\
$l$& Length of the leaf\\
$w$& Width of the lamina\\
$t$ & Thickness of the lamina\\
$d$ & Diameter of the petiole/midvein\\
$g$ & Acceleration of gravity \\
$\rho$ & Leaf mean density \\
$G$ & Gravity loading per unit of length\\
$C_0$& Palm curvature without external loading\\
$C$ & Palm curvature with external loading\\
$l_x$ & Horizontal projection of palm length\\ 
$E$ & Young modulus of the leaf\\
$I$ & Second moment of area of the leaf\\
$\mathcal{E}$ & Daily collected solar energy\\
$\phi$ & Solar inclination angle\\
$\lambda$ & Lagrange multiplier\\
\hline
\end{tabular}\end{center}
\caption{Table of parameters}
\end{table}



\section{Introduction}\label{introduction}

Leaves are responsible for intercepting light and creating sugars from photosynthesis \citep{Life_leaf}. The ability of a plant to produce biomass, to grow and reproduce (in other words to compete for survival) depends on the efficiency of photosynthesis. Photosynthesis yield depends on various parameters, mainly gas exchange, leaf temperature and light interception \citep{Farquhar1980}. Different factors may have a role on those key parameters: leaf perspiration, stomata aperture and ability to flutter may alter heat and gas exchange at the level of the leaf \citep{Roden1993, Roden2003}. 

Leaf orientation towards the sun's rays plays a key role for light interception \citep{Tadrist2014} and different strategies have been adopted by the vegetal kingdom. In the first strategy, leaf orientation has no preferred direction to collect diffuse light that comes from every direction. In the second strategy, leaves have to be properly oriented to collect direct sun light. More complex strategies to improve the rate of photosynthesis  also exist, giving rise to rather dynamical plant behaviour depending on environmental parameters. One example is the time-dependent orientation of leaves to follow the sun's position in the sky, but they can be even more complex. For instance, during a drought period, the inability to access water from the soil prevents the plant from perspiring. The leaf-refreshing effect of transpiration is cancelled, leaf temperature increases and the rate of photosynthesis drop to zero. To avoid such dramatic loss, leaves' orientations are changed to intercept less light, reduce leaf temperature and keep leaves active for photosynthesis, \citep{Life_leaf, GonzalezRodriguez2015}.

\begin{figure}\begin{center}
\begin{psfrags}\psfrag{a}[c][c]{(a)}\psfrag{b}[c][c]{(b)}\psfrag{c}[c][c]{(c)}\psfrag{i}[c][c]{i}\psfrag{ii}[c][c]{ii}\psfrag{iii}[c][c]{iii}\psfrag{iv}[c][c]{iv}\psfrag{0}[c][c]{0}\psfrag{p2}[c][c]{$\pi/2$}\psfrag{X100}[c][c]{(X100)}\psfrag{4m}[c][c]{4 m}
\includegraphics[width=0.4\textwidth]{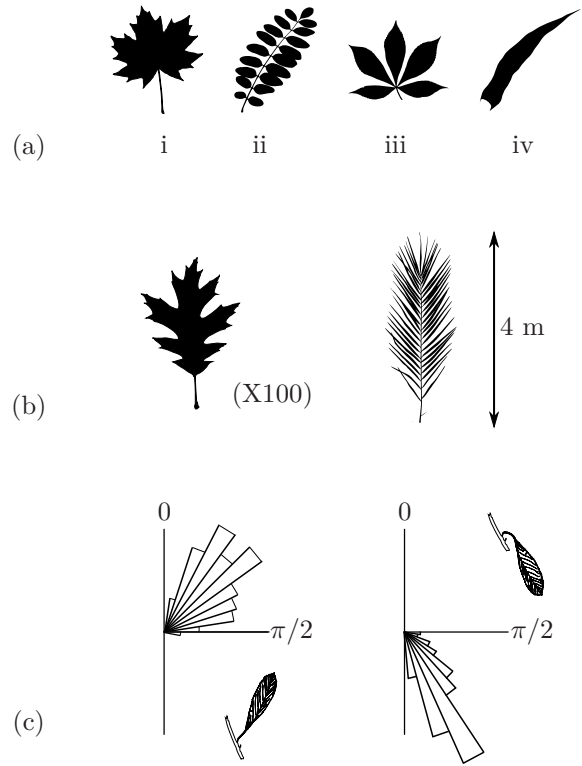}
\end{psfrags}
\caption{(a) Variety of shapes of leaves: (i) simple leaf, (ii) palmate leaf, (iii) pinnate leaf. (iv) sessile leaf, without petiole. (b) Variety of sizes: the first leaf is an oak leaf which has typical dimensions on the order of few centimetres, compared with phoenix palm that has a typical dimension of a few meters. In proportion, the oak leaf size has been magnified by a factor 100. The largest leaf was measured on \textit{Raphia regalis} with a length of 25.11 m \cite{Halle2005}. (c) Variety of leaf orientations in term of leaf inclination angle distribution. The first distribution depicts erected leaves whereas for the second one (\textit{Trema aspera}) most of the leaves are hanging down.}
\label{Figure1}
\end{center}\end{figure}

Because of these reasons, one may think that leaf orientation, leaf shape and leaf size would be parameters subjected to strong pressure of selection because of their role in light interception.
The design of the leaf (thickness, length, width, petiole, midvein, secondary veins, etc.) must be optimal to obtain the larger rates of photosynthesis without threatening the plant's life. The highest photosynthetic rates of plants can reach up to 30\% with a mean rate around 3\% \citep{Biologie_vegetale}. In this context, it is surprising to observe such a variability in leaf shape, sizes and orientations (see Fig. \ref{Figure1}). Discussions about plant mechanical optimality make a long story. The first authors to introduce the concept of plant design constrained  by mechanics were \citet{McMahon1976}, who proposed that tree height and tree width are bound features. A few discussions of plant optimality concern branches \citep{wei2012branch} and leaves \citep{Niklas1992, Niklas1993, Plant_physics}, and \citet{Jensen2013} have shown that leaf size is limited by optimal sap flow in tall trees.
 
Simple calculation of light interception with geometrical arguments shows that the optimal position to collect direct sun light is orthogonal to the local gravity field; see \ref{annex_light} and \citet{Tadrist2014}. On terrestrial plants, leaves grow on branches and must support their own weight in order not to hang down. Some bio-material used to make the leaf must be dedicated to create surface area for light interception, but the rest of the bio-material should be used for mechanical self-support. On the one hand, if too little bio-material is used for self-support, the leaf will hang down and despite the large amount of bio-material used to create surface area, it will not intercept much light. On the other hand, if too much bio-material is used for self-support, the leaf will be properly oriented to intercept light but will have no surface area to collect light. An optimal mass allocation trade-off exists between those two extrema.  
  What would be the shape of a leaf that has optimally allocated biomass? Note that when no self-support is needed, all the biomass would be used to create surface area for light interception. This is the case for water lilies that occupy the interface between air and water. 

Answering questions considering optimality in nature is not easy because of the large number of functions performed by an organ and because of the still larger number of environmental parameters to take into account. Those functions may naturally encourage antagonistic shapes; for instance, the optimal leaf would be thin enough to have a large surface area-to-volume ratio to enhance gas and heat exchange, but also thick enough for efficient transport of water and sap in xylem and phloem. Abiotic and biotic stresses are also shaping factors for the leaf. For example, a leaf may be designed to flutter to increase photosynthesis rate \citep{Roden2003}, to expel water drops and prevent fungus attacks or simply to remove hervibory insects \citep{Yamazaki2011}. Leaves have also developed mechanical tricks to be stiffer for the same amount of bio-material and thus to reduce the amount of biomass needed to support their own weight. Those tricks are (i) the inhomogeneity of the leaf tissues (Xylem and phloem vessels are much more lignified -and thus stiffer- tissues than mesenchymatous cells.), (ii) the anisotropic placement of the tissues and (iii) the shape of the leaf itself. For instance, the shape of the leaf could lead to a stiffer U-shaped petiole \citep{Ennos2000} or stiffer lamina through curvature-induced rigidity \citep{Barois2014}.

In this paper, we focus on the trade-off between self-support and creation of surface area for light interception. Our approach is based on a basic mechanical modelling of the leaf which neglects the different stresses or parameters of selection that apply on leaves, nor on the complex mechanical tricks developed to enhance leaf rigidity while minimizing biomass use.

Optimal leaf shape is examined for the giant monocots leaves, especially palms of palm trees and banana trees. Palm trees belong to the large family of \textit{Arecaceae} (more than 2600 species) within the monocots clade. In this family, plants exhibit leaves of different sizes and different shapes. We choose here to study plants with consistent, simple leaf geometry: one short petiole and one long lamina with one major vein. In the first part of this article we take advantage of this simple geometry to describe theoretically what would be the optimal leaf of a palm tree. In the second part of this article, we describe the measurements done on actual palm trees and banana trees. Finally, we compare the theoretical results with the measurements on monocot trees. We show that the shape of larger leaves is close to the predicted optimal shape whereas the smallest palms shapes differ strongly from prediction. We predict a minimal size for which our model applies. For smaller leaves, self-support does not appear to be the strongest factor of selection.

\section{Model}\label{model}

We aim at modelling what would be the optimal shape of a leaf under mechanical self-support constraints. We develop here a simple model based on mechanical considerations. For the sake of clarity, mechanical and geometrical assumptions are made for the considered leaf.

\subsection{Model assumptions} \label{model_assumptions}

\begin{figure*}\begin{center}
\begin{psfrags}
\psfrag{l}[c][c]{$l$}\psfrag{w}[c][c]{$2 w$}\psfrag{a}[c][c]{(a)}\psfrag{b}[c][c]{(b)}\psfrag{c}[c][c]{(c)}\psfrag{d}[c][c]{(d)}\psfrag{r1}[c][c]{1}\psfrag{r2}[c][c]{2}\psfrag{l}[c][c]{l}\psfrag{dd}[c][c]{$d$}\psfrag{t}[c][c]{$t$}\psfrag{th}[c][c]{$\theta$} \psfrag{cross section}[c][c]{cross section}
\includegraphics[width=\textwidth]{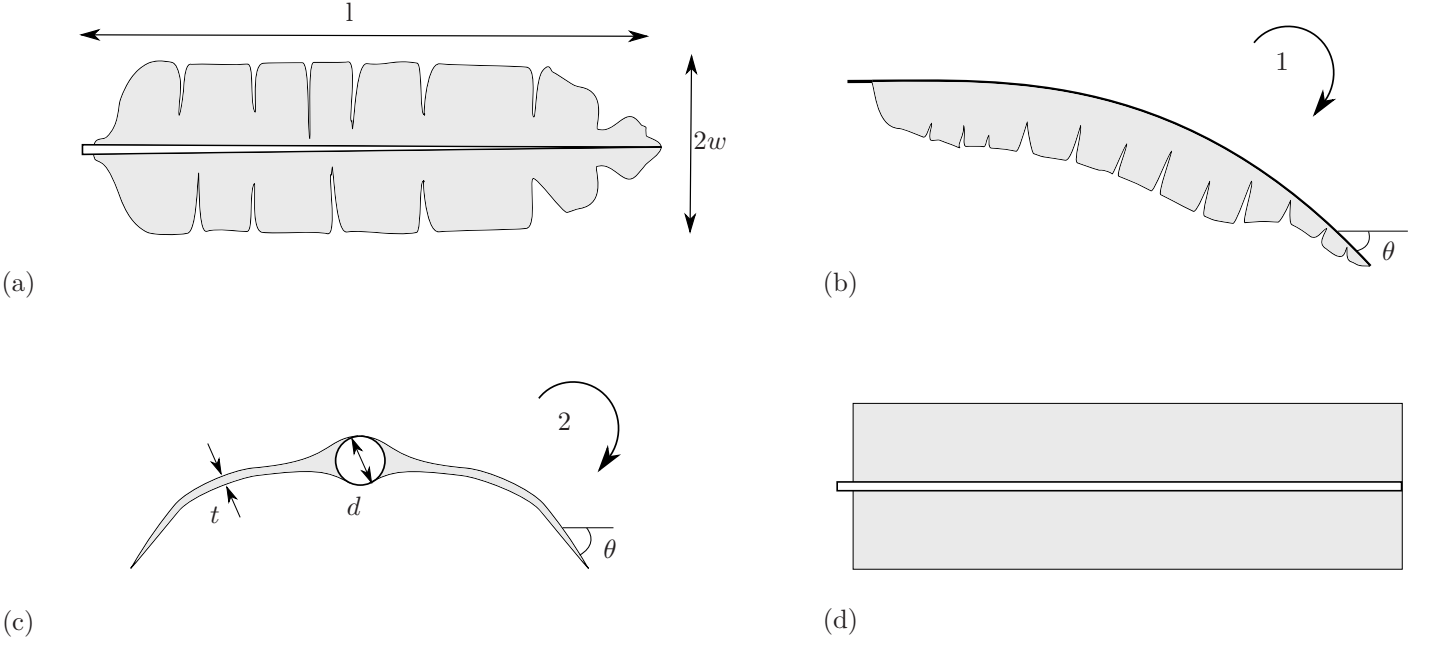}
\end{psfrags}
\caption{Schematic views of a giant monocot leaf. (a) top view. (b) lateral view. The leaf hangs down because of its own weight. $\theta$ is the largest inclination angle of the leaf. This bending direction is noted 1. (c) Cross section view of the leaf. Again, the leaf hangs down because of its own weight and $\theta$ is the largest inclination angle of the leaf in this bending direction. This bending direction is noted 2. (d) Top view of the idealized leaf used for the model.}
\label{Figure2}
\end{center}\end{figure*}

We detail the assumptions as follow: first, the geometry of the considered leaf is chosen as the geometry of a palm of \textit{Phoenix Canariensis}. Such a palm is composed of a short petiole that quickly becomes the midvein of the palm. The diameter of the midvein at the base of the lamina is denoted $d$. The lamina has a length $l$, a width $w$ and a thickness $t$. The leaf geometrical properties are sketched in Fig. \ref{Figure2}. Second, we consider that all the bio-material used to create the leaf is isotropic, has the same Young modulus $E$ and the same density $\rho$. Its production needs the same amount of bio-energy. Under this strong assumption, we neglect the inhomogeneity of the different tissues that comprise the leaf (xylem/phloem vessels, mesenchymatous cells, cuticle, etc.) Those tissues have non-similar mechanical properties. This will be discussed in the last part of the paper as well as the growth process.
Third, we also neglect all mechanical tricks to strengthen the midvein such as U-shaped midvein \citep{Ennos2000}, turgidity-dependent rigidity \citep{Nilson1958, Faisal2010} and curvature induced rigidity of the lamina \citep{Barois2014}.
Fourth, for the sake of simplicity, we consider that the petiole is clamped perpendicular to the gravity. Finally, we assume that the midvein weight is negligible compared to the lamina weight. This assumption can be easily released and optimal shapes can be computed numerically but it does not permit to access analytical solutions any more.  

\subsection{Optimal leaf}

We consider a leaf supporting its own weight $M$ within a gravitational acceleration $g$ for a given volume of biomass $V$ available to build the structure. The leaf bends in the two directions indicated in Fig. \ref{Figure2}b and \ref{Figure2}c. The first direction is along the midvein and the second is perpendicular to the midvein. We note $\theta$ the final angle of the leaf. The maximal angle of a beam horizontally clamped at its one end under constant gravity loading $G$ is 
\begin{equation}
\theta = \frac{G L^3}{6EI}\, , 
\label{eq:1}
\end{equation}
where $I$ is the second momentum of area ($I= \pi d^4/64$ for a circular beam in bending). This equation is adapted for bending in direction 1 
\begin{equation}
\frac{2\rho\, g\, w\, t\, l^3}{6 E\, (\pi\, d^4/64)} =\theta \,,
\label{eq:X}
\end{equation}
and in direction 2
\begin{equation}
\frac{\rho\, g\,l\,t\, w^3}{6 E\, ( t^3\,l/12)} =\theta \,.
\label{eq:Y}
\end{equation}

These two equations are the only mechanics-derived expressions in this model. Volume, $V$, and surface area, $S$, of the leaf can be represented by
\begin{equation}
 V = 2 w\, t\,l + \pi\, l\, d^2/4 \qquad \mathrm{and} \qquad S=l\,w \,.
\label{eq:vol}
\end{equation}

Maximization of the surface of the leaf $S$ under the constraint of fixed volume of biomass $V$ and adequacy with mechanical self-support, Eq. (\ref{eq:X}) and (\ref{eq:Y}), provides four equations that give $l$, $w$, $d$ and $t$. Calculation details are presented in the \ref{annex_biomass}. Finally, we obtain
\begin{equation}
l= \left(\frac{2\,3^2\, \theta}{5\,11\,\pi}\right)^{1/4}\left(\frac{E}{\rho g}\right)^{1/4} V^{1/4} 
\label{eq:scale_l}
\end{equation}

\begin{equation}
w= \left(\frac{5^5\, \pi\, \theta}{2^7\,3^{3}\, 11^{3}}\right)^{1/10}\,\left(\frac{E}{\rho g}\right)^{1/10} V^{3/10} 
\label{eq:scale_w}
\end{equation}

\begin{equation}
t= \left(\frac{\pi^3\, 5^{15}}{3^{9} \,2^{11} \,11^9\, \theta^7}\right)^{1/20}\left(\frac{E}{\rho g}\right)^{-7/20} V^{9/20} 
\label{eq:scale_t}
\end{equation}

\begin{equation}
d= \left(\frac{2^{11}\,3\,5}{11^3 \,\pi^3 \,\theta}\right)^{1/8}\left(\frac{E}{\rho g}\right)^{-1/8} V^{3/8} 
\label{eq:scale_d}
\end{equation}

Considering different leaf geometry, with a cone-shaped midvein for instance (see \citet{wei2012branch}) would lead to similar results but with slightly different pre-factors. To bear in mind some numerical values, a giant leaf with a volume $V=1.0 \times 10^{-3}$ $\mathrm{m}^{3}$, a typical Young's modulus of $E=1.0 \times 10^8$ Pa and a density of $\rho=1.3 \times10^3$ kg.$\mathrm{m}^{-3}$ the model gives a length of 80 cm, a width of 16 cm, a midvein diameter of 2.8 cm and a thickness of 0.9 mm.

\section{Material and Methods}

To assess the hypothesis that leaves are optimally designed for self-support applies in nature, we measured leaf characteristics of palm trees and banana trees. 

\subsection{Samples} \label{palmtrees}

The palms of various species of monocots have been collected in the garden of the Rayol domain (N $43^\circ\,9'\,22''$ E $6^\circ\,28'\,54''$) and in private gardens in La Ciotat (N $43^\circ\,10'\,25''$ E $5^\circ\,36'\,18''$) and in \'Eguilles (N $43^\circ\,32'\,39''$ E $5^\circ\,20'\,22''$) all located in the south of France. The species and the characteristics of the palms used in our experiments are listed in Table \ref{tab:palms}. The palms selected for the study were cut from palm trees of different species living in different habitats and thus have different characteristics. Some are drought resistant such as \textit{Sica} whereas some require wet environment like the ferns \textit{Dicksonia Antartica} and \textit{Cyathea Cooperi}. Most of the species chosen in the study were grown in south of France but originate from all over the world (south America: \textit{Butia}, \textit{Syagrus Romanzoffiana}; Asia: \textit{Banana tree}, \textit{Sica}; Africa: \textit{Phoenix Canariensis}; Oceania: \textit{Rhopalostylis Sapida}, \textit{Cyathea Cooperi}, \textit{Macrozamia Communis}.)

\begin{figure*}[b]
\begin{center}
\begin{tabular}{| p{0.6cm} | p{3.7cm} | p{1.5 cm} | *{6}{p{0.7cm}|} p{1.1cm}| p{1.6cm}| p{1cm}| }
\hline
Exp. & Species & Location & $l$ (cm) & $2 \, w$ (cm) & $d$ (mm) & $t$ ($\mu$m)  & $V$ (cm$\,^3$) & $M$ (g) & $\rho$ (g/cm$\,^3$) & $E$ (Pa) \\
\hline
1 & \textit{Banana tree} & \'Eguilles & 49 & 12.2 & 5.5 & 210 & 21 & 13 & 0.62 &$8.3 \times 10^6$  \\
2 & \textit{Sica} & \'Eguilles & 84 & 9.2 & 11.2 & 720 & 95 & 91 & 0.96 &$4.5 \times 10^{10}$  \\
3 & \textit{Phoenix} & \'Eguilles & 124 & 17.9 & 20.1 & 438 & 315 & 332 & 1.05 & $2.7 \times 10^8$  \\
4 & \textit{Macromazia Communis} & Rayol & 134 & 26.3 & 9.0 & 591 & 155 & 151 & 0.97 & $2.8 \times 10^9$ \\
5 & \textit{Cyathea Cooperi} & Rayol & 167 & 69 & 12.9 & 147 & 390 & 311 & 0.78 & $2.1 \times 10^9$ \\
6 & \textit{Buttia} & \'Eguilles & 190 & 21.2 & 10.9 & 427 & 340 & 385 & 1.13 &  $1.3 \times 10^{10}$  \\
7 & \textit{Buttia} & Rayol & 205 & 58.7 & 15.7 & 481 & 1303 & 1473 & 1.13 & $4.8 \times 10^9$ \\
8 & \textit{Dicksonia Antartica} & Rayol & 241 & 42.1 & 13.3 & 61 & 400 & 631 & 1.58 & $3.0 \times 10^9$  \\
9 & \textit{Syagrus Romanzoffiana} & Rayol & 270 & 91.7 & 27.3 & 201 & 1717 & 1796 & 1.04 & $1.4 \times 10^9$\\
10 & \textit{Phoenix} & La Ciotat & 300 & 73.8 & 19.7 & 310 & 840 & 946 & 1.12 & $6.4 \times 10^8$ \\
11 & \textit{Rhopalostylis Sapida} & Rayol & 378 & 100 & 31.2 & 310 & 2280 &  2434 & 1.07 & $5.1 \times 10^8$ \\
\hline
\end{tabular}
\end{center}
\caption{Characteristics of palms studied experimentally: Species [column 1]; Tree location [column 2]; Length of the palm $l$ [column 3]; Mean width of the palm $2w$ [column 4]; Diameter of the petiole at it base $d$ [column 5]; Mean thickness of the palm $t$ [column 6]; Volume of the palm $V$ [column 7]; Mass of the palm $M$ [column 8]; density of the palm $\rho=M/V$ [column 9]; Young's modulus of the palm $E$ [column 10].}
\label{tab:palms}
\end{figure*}


\subsection{Measurements}
\begin{figure}\begin{center}
\begin{psfrags}
\psfrag{a}[c][c]{(a)}\psfrag{b}[c][c]{(b)}\psfrag{m}[c][c]{$M_e$}\psfrag{p}[c][c]{$M_e\,g\,l_x$  (kg.$\mathrm{m}^2/\mathrm{s}^{-2}$)}\psfrag{c}[c][c][1][90]{$C-C_0$ ($\mathrm{m}^{-1}$)}\psfrag{0}[c][c][0.8]{0}\psfrag{0.1}[c][c][0.8]{0.1}\psfrag{0.2}[c][c][0.8]{0.2}\psfrag{0.3}[c][c][0.8]{0.3}\psfrag{4}[c][c][0.8]{4}\psfrag{8}[c][c][0.8]{8}\psfrag{12}[c][c][0.8]{12}
\includegraphics[width=0.45\textwidth]{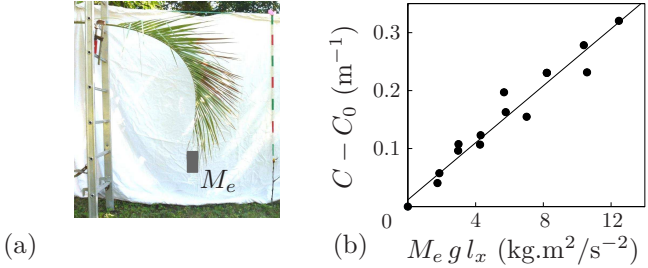}
\end{psfrags}
\caption{(a) Palm of a Buttia palm tree (Exp. 2) clamped at its base. A mass of $M_e=0.4$ kg is attached to its free end. (b) Plot of the relative curvature $C - C_0$ of the base of a palm of a Syagrus Romanzoffiana tree, see experiment number 10 in Table \ref{tab:palms}, as a function of the torque $M g l_x$ imposed at the free end. The slope gives an estimation of $EI$ in $\mathrm{kg.m}^3/\mathrm{s}^2$.}
\label{fig:method}
\end{center}\end{figure}

We have measured geometrical and mechanical parameters of the leaves, their dimensions, $l$, $w$, $d$ and $t$, their volume $V$, their mass $M$, their density $\rho$ and their Young's modulus $E$.
Immediately after being cut from the tree, fresh palms were clamped horizontally from their bases and tested with extra masses $M_e$ attached to their free end. We took side pictures of this system, as shown in Fig. \ref{fig:method}a, in order to estimate the Young's modulus $E$ of the palms. After that, all the dimensions of the palm (length $l$, width $w$, petiole diameter $d$ and thickness $t$) were measured. Palms used in our experiments vary from 48 to 378 cm in length and from 12 to 100 cm in width. Also, we determined the total mass $M$ of the palm with a precision scale and reported a variation between 13 and 2434 g for the considered palms. Finally, the volume $V$ of the palm has been estimated by measuring its displacement of water when submerged in a tank. In our experiments, the volume of a palm varied between 21 to 2280 $\rm{cm^3}$. The measurements made on palms of different species are gathered in Table \ref{tab:palms}. 

\section{Results} 

\subsection{Density}

The deduction of the density $\rho=M/V$ of palms are listed in column 10 of Table \ref{tab:palms}. The density of palms considered in our experiments varies from $0.78\times10^{3}$ to $1.58\times10^{3}\,\mathrm{kg.m}^{-3}$. This observation means that palms of various species of monocots are denser than water and must sink. Such a counter-intuitive prediction for vegetable materials has been verified in a pool.

\subsection{Young's modulus}

The Young's moduli $E$ of the different palms considered in our experiments have been deduced from their shape when clamped with an extra-load $M_e\, g$ applied to their free end, as shown in Fig. \ref{fig:method}a. In such a situation, the extra load imposes a torque at the base of the palm which is expressed as $M_e\,g\, l_x$, where $l_x$ is the length of the palm projected on the horizontal direction. Assuming that the palm behaves as an elastic beam, the previous torque is compensated by the elastic torque $EI (C - C_0)$ at its base, where $C$ is the curvature of the palm just adjacent to the clamp and $C_0$ the natural curvature of the palm at the same position when $M_e=0$. Finally the torque equilibrium at the base of the palm provides

\begin{equation}
EI (C- C_0) = M_e\, g \,l_x \,.
\end{equation}

Experimentally, we measured the two quantities $C-C_0$ and $l_x$ for various extra masses $M_e$ with image analysis. Figure \ref{fig:method}b shows the relation between the relative curvature $C-C_0$ at the base of a palm of a Syagrus Romanzoffiana as a function of the torque $M_e\, g\, l_x$ applied to its free end. The correlation of the two quantities provides an estimation of the flexural rigidity $EI$ of the palm along its principal direction. Assuming that in this situation, the second moment of area $I$ is the one of a cylindrical beam of diameter $d$, ($I=\pi d^4 /64$), we estimate the Young's modulus $E$ of the palm. The same procedure has been done for every palm and the estimations of Young's modulus are reported in Table \ref{tab:palms}. For palms considered in this study, $E$ is in a range from $ 10^7$ to $10^{10}$ Pa consistently with \cite{Niklas_allometry}.

\section{Comparison model \& measurements and Discussion}

\subsection{Comparison}

\begin{figure*}\begin{center}
\begin{psfrags}
\psfrag{a}[c][c]{(a)}\psfrag{b}[c][c]{(b)}\psfrag{c}[c][c]{(c)}\psfrag{d}[c][c]{(d)}
\psfrag{w}[c][c][1][90]{$w$ (cm)}\psfrag{l}[c][c][1][90]{$l$ (m)}\psfrag{t}[c][c][1][90]{$t$ (mm)}\psfrag{dd}[c][c][1][90]{$d$ (cm)}
\psfrag{0}[c][c]{0}\psfrag{1}[c][c]{1}\psfrag{2}[c][c]{2}\psfrag{3}[c][c]{3}\psfrag{4}[c][c]{4}\psfrag{5}[c][c]{5}\psfrag{6}[c][c]{6}\psfrag{0.5}[c][c]{0.5}\psfrag{1.5}[c][c]{1.5}\psfrag{20}[c][c]{20}\psfrag{40}[c][c]{40}\psfrag{60}[c][c]{60}\psfrag{80}[c][c]{80}
\psfrag{vt}[c][c]{$\left( E / \rho g \right)^{-7/20} V^{9/20}$ (mm)}\psfrag{vl}[c][c]{$\left( E / \rho g \right)^{1/4} V^{1/4}$ (m)}\psfrag{Vd}[c][c]{$\left( E / \rho g \right)^{-1/8} V^{3/8}$ (cm)}\psfrag{Vw}[c][c]{$\left( E / \rho g \right)^{1/10} V^{3/10}$ (cm)}
\includegraphics[width=0.8\textwidth]{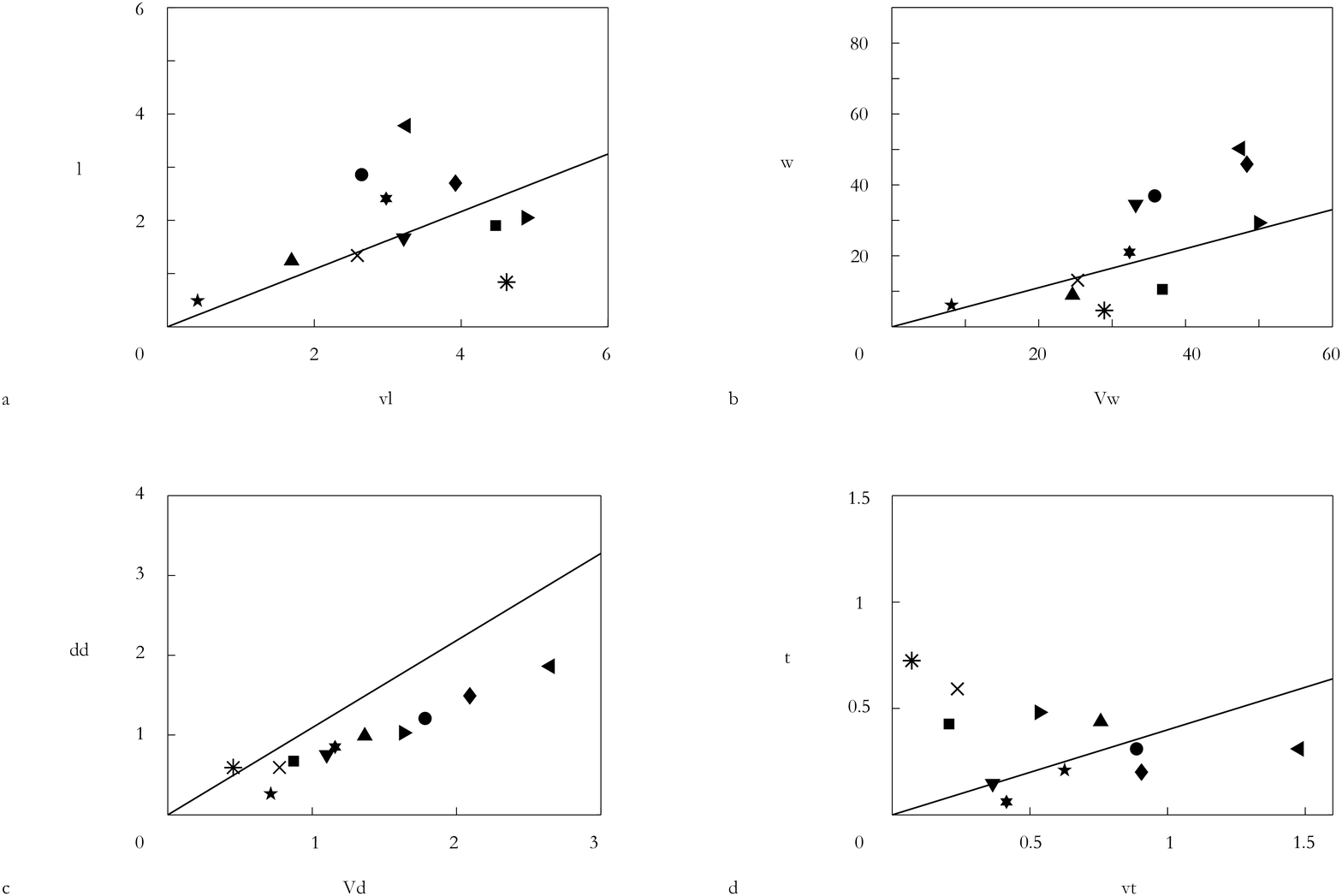}
\end{psfrags}
\caption{(a) Palm length $l$ of various species of palm tree as a function of $\left( E / \rho g \right)^{1/4} V^{1/4}$. $\bigstar$ \textit{Banana tree} (Exp. 1),
\ding{83} \textit{Sica} (Exp. 2),
$\blacktriangle$ \textit{Phoenix Canariensis} (Exp. 3),
$\times$ \textit{Macrozamia Communis} (Exp. 4),
$\blacktriangledown$ \textit{Cyathea Cooperi} (Exp. 5),
$\blacksquare$ \textit{Buttia} (Exp. 6),
$ \blacktriangleright$ \textit{Buttia} (Exp. 7),
\ding{88} \textit{Dicksonia Antartica} (Exp. 8),
$\blacklozenge$ \textit{Syagrus Romanzoffiana} (Exp. 9),
\ding{108} \textit{Phoenix Canariensis} (Exp. 10),  
$ \blacktriangleleft$ \textit{Rhopalostylis Sapida} (Exp. 11).
(b) Mean width $w$ of various palms as a function of $\left( E / \rho g \right)^{1/10} V^{3/10}$. (c) Mean thickness $t$ of palms of different species as a function of $\left( E / \rho g \right)^{-7/20} V^{9/20}$. (d) Diameter of the petiole $d$ at the base of palms of different species as a function of $\left( E / \rho g \right)^{-1/8} V^{3/8}$. The black solid lines represent respectively Eq. (\ref{eq:scale_l}), (\ref{eq:scale_w}), (\ref{eq:scale_t}) and (\ref{eq:scale_d}) for $\theta=20^\circ$.
}
\label{fig:scale}
\end{center}
\end{figure*}

In order to compare the experimental data with the model developed previously, we plot the palm length, mean width, mean thickness and the mean petiole diameter as a function of the scaling laws predicted by equations (\ref{eq:scale_l}), (\ref{eq:scale_w}), (\ref{eq:scale_t}) and (\ref{eq:scale_d}) in Fig. \ref{fig:scale}.

First, for each parameter the model predicts the actual value within an order of magnitude. Second, the evolution of the data with the scaling laws are consistent. Third, the prefactors of the model yield an estimation of the parameter close to their absolute value. 

One may notice that the \textit{Sica} leaf does not follow the general trend for any of the parameters. Such a leaf is not optimally designed for self-support consistently with our observation; gravity does not impact the shape of a mature \textit{Sica} leaf. As a general rule, when a leaf does not change of shape when gravity is reversed (Moulia's test\footnote{From Bruno Moulia, INRA Clermont-Ferrand.}), for instance when its curvature does not change when flipped upside down, the leaf is too stiff and not optimally designed for self-support. To build such a leaf, too much biomass was used for self-support and one may think that others factors of selection prevail. 

\subsection{Discussion}

The agreement between experiment and prediction supports our hypothesis that leaves are optimally designed for self-support. The discrepancy in terms of absolute value may come from the simplification of the geometry of the leaf made in the model (see section \ref{model_assumptions}). For all the leaves considered here, the midvein is not cylindrical but cone-shaped. Also, the non-uniform elasticity of the leaf can explain the difference between measurements and theory. Finally, the mechanical tricks make the leaf stiffer with the same amount of bio-material and thus must increase the length and the width of the leaf. We expected that our predictions of $l$ and $w$ would be lower than the measured values, which is the case. Note that palms are composed by separated leaflets which reduce the apparent density and increase $w$ and $l$, see Eq. (\ref{eq:scale_l}) and (\ref{eq:scale_w}).

In contrast, the model predicts the leaves' thickness for most of the leaves presented here but does not predict accurately the measurements for the smallest \textit{Sica} leaf, which is too stiff compared to the optimal self-support prediction.

There is a minimal size of leaves for which the proposed model applies. If we consider that a leaf cannot be thinner than 100 $\mu$m, we obtain a critical leaf length of 23 cm and a critical width of 4 cm (for $E= 10^8$ Pa, $\rho=1.3\times 10^3$ kg.$\mathrm{m}^{-3}$ and $\theta=20^\circ$). This means that small leaves encountered in nature are too stiff and thus not optimized for self-support. Other selection parameters may influence on these leaves to determine their shape. Among these constraints, two example are wind resistance and resistance to tearing. Also, \cite{Jensen2013} discussed the maximal size of leaves resulting from the optimality of sap flow efficiency.

Our study shows that some large leaves are subjected to biomass allocation optimization. But, one may ask if it is preferable to concentrate much of the biomass of the tree in one leaf or if it is more efficient to create a large bunch of small leaves. According to our approach, there is no optimal position on this point. Indeed, the volume of the leaf dedicated to light interception is $V_{\mathrm{light}}=2 l\,w\,t$ that scales as $V_{\mathrm{light}}\propto V^{1/4}\,V^{3/10}\,V^{9/20}\propto V$. In proportion, the amount of leaf useful for light interception is independent of the size of the leaf; having a bunch of small leaves or only one huge leaf of the same volume gives the same results in terms of light interception. In reality, the repartition between surface and thickness is different. A bunch of small leaves will have a larger surface to intercept light but a smaller thickness. This indicates that the transmission coefficient of the small leaves should be larger than the one of large leaves. 

This particularity of leaf shape might lead to different light interception strategies:
(i) In a strong light environment, a thick lamina allows a leaf to intercept a lot of light directly (light interception is exponential with lamina thickness), whereas a thin lamina will intercept less light and thus transmit more to under canopy leaves. One large leaf is favourable in this light environment.
(ii) In a low light environment, a thin leaf already intercepts all of the available light. Being thicker is no longer efficient to intercept more light. However, with thin leaves the potential surface area of light interception is larger than with thick leaves. A bunch of small leaves is beneficial in such an environment.

Furthermore, areas with a strongly lit environment are located close to the inter-tropical region, where the light is mainly direct light and thus the sun-leaf orientation matters. At higher latitudes, the available light is lower intensity and diffuse light that does not come from a preferred direction - leaf orientation does not matter. This reinforces our earlier point: in the inter-tropical region, leaves have to deal with strong direct light, so building one leaf properly oriented (\textit{i.e.} horizontal) is favourable whereas in higher latitude regions, building a bunch of small leaves oriented in different direction would be more efficient in low and diffuse light environment. 
Because of these points, it may not be surprising that large palm trees are found in inter-tropical regions rather than close to the poles.

This hypothesis that leaf size is repartitioned with Earth's latitude needs to be verified in a statistical manner. Many selection factors may influence the leaf shape or change the optimal leaf size; see for instance the paper by \citet{Jensen2013}.  

Our analysis has many biases on which we comment in the following: (i) There are many shapes of leaves in nature. We believe that for large leaves, arguments of optimal biomass usage would lead to similar results regardless the actual shape of the leaf. (ii) The leaf is created with a growth process that tends to change the mechanical properties of the leaf (pre-stresses, inhomogeneities, natural curvatures). We believe that the arguments presented here apply regardless of the mechanical specificity of each leaf. This is because the dependence on global shape is much stronger than the dependence on any mechanical trick (see exponents in Eq. (\ref{eq:scale_l}), (\ref{eq:scale_w}), (\ref{eq:scale_t}) and (\ref{eq:scale_d}), always smaller than 1/2).
The leaves which measured dimensions are strongly different from predictions of the model would be leaves with a different strategy of orientation for light collection (such as diffuse light collection) or fulfilling other selection parameters.

\section{Conclusion}

This study shows that large leaves, as the palms studied here, are optimized to support themselves for direct light interception. However, most of the leaves of terrestrial plants are oversized for self-support and thus not optimal. Leaf over-sizing is easily observable: when you take a small leaf in your hand and flip it regarding the gravity, nothing happens and the leaf keeps its own curvature which results from growth (Moulia's test). This over-sizing supports the idea that different selection pressures are more relevant to explain the shape of the leaf rather than the one exposed in this paper. Among these selection pressures are storm survival, drought resistance, herbivory resistance, and enhancement of heat and gas exchange. The large number of selection pressures indicates that leaves are multi-functional organs for which the shape is the result of a complex trade-off. However, depending on leaf fitness, environment, and life history, one among those selection pressures may play a stronger role into leaf shaping than the others. The determination of which selective pressures shape the leaf in a given environment is a fascinating question that still requires further investigation. 
A better understanding of how competing selective pressure favours some leaves shapes has strong agricultural implications, e.g. the artificial selection of plants to adapt to non-native habitats.

Another perspective of this work could be to relax some assumptions made in the model. Particularly, the simple geometry of palm leaves assumed in the model can be extended to other geometries using an evolutionary algorithm. We expect that such a procedure would predict the optimal geometry of a large leaf with more details (the midvein geometry, lamina shape and even the presence of secondary veins). 

\vspace{0.3cm}

\section*{Acknowledgments}

The authors gratefully acknowledge the direction of Rayol's domain for providing us most of the samples used in this study. We warmly thank Karen Texier, Alexandre Lanza and Cyrielle Decl\`eve who have preferred to cut palms rather than lounging under their shade or simply sunbathing! We also thanks Kevin Eckes for careful review of the manuscript. The authors are thankful to St\'ephane Dorbolo, Emmanuel de Langre, Bruno Moulia, Alain Menseau and Marc Saudreau for fruitful discussions. Moreover, a special thank goes to St\'ephane Dorbolo for his contagious passion of the vegetal kingdom which has motivated this work.

\bibliographystyle{model2-names}

\appendix
\section{Light interception}\label{annex_light}

\begin{figure}[htp!]
\begin{center}
\begin{psfrags}
\psfrag{t}[c][c]{$\theta$}\psfrag{p}[c][c]{$\phi$}
\includegraphics[height=3cm]{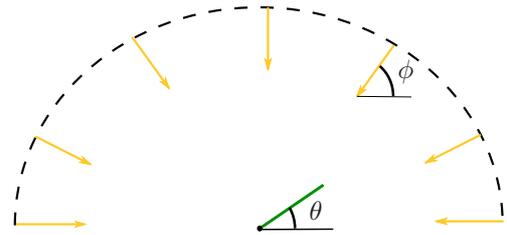}
\end{psfrags}
\caption{Notation used to describe light interception by a leaf.}
\label{fig:sketch}
\end{center}
\end{figure}

The goal of this appendix is to explain why a horizontal and flat position maximizes sunlight interception. Our approach considers a leaf as a thin plate, inclined by an angle $\theta$  with the local gravity, Fig. \ref{fig:sketch}. A simple sun path across the sky is indicated by the dotted line. We assume the sun provides a constant solar energy whatever its position on the sky (This is a simple assumption but transmitted solar energy is larger when the sun is at its zenith. This effect strongly favours the horizontal position). Moreover we suppose that the sun shines from each position parallel sunbeams with an angle $\phi$ with respect to the horizontal. For the exact and detailed calculation of light interception, see \citet{Pisek2011} and \citet{Tadrist2014}.

The total solar energy collected by the leaf on one day $\mathcal{E}$ is the sum of every sunbeam ($0<\phi<\pi$)

\begin{equation}
\mathcal{E}\propto \left(\int_\theta ^\pi \sin(\phi - \theta) \, d \phi + \int_0 ^{\pi + \theta} \sin(\phi - \theta) \, d \phi\right)
\label{eq:totalenergy}
\end{equation}
The integration over $\phi$ yields

\begin{equation}
\mathcal{E}\propto \left( 1 + \cos \theta \right)
\label{eq:totalenergy2}
\end{equation}

One observes that $\mathcal{E}$ is maximum when $\theta(s)=0$. The optimal orientation of a leaf for sunlight interception is a horizontal. Taking into account the variation of sunlight power during the day (\textit{i.e.} with respect to solar inclination, $\phi$) would reinforce the previous result.

\section{Optimization of biomass allocation}\label{annex_biomass}

We develop here the calculations to derive Eq. (\ref{eq:scale_l}), (\ref{eq:scale_w}), (\ref{eq:scale_t}) and (\ref{eq:scale_d}) from the three first equations Eq. (\ref{eq:X}), (\ref{eq:Y}), (\ref{eq:vol}) and surface optimization.

With Eq. (\ref{eq:X}), $t$ writes
\begin{equation}
t=\sqrt{2} \left(\frac{E\theta}{\rho g}\right)^{-1/2} w^{3/2}\,.
\label{eq:int1}
\end{equation}

To simplify notations, in the following, the length ${E\theta}/{\rho g}$ is noted $\Theta$. With this notation, $t$ reads
\begin{equation}
t=\sqrt{2} \Theta^{-1/2} w^{3/2}\,.
\label{eq:int1_bis}
\end{equation}

With Eq. (\ref{eq:Y}), $d^4$ is
\begin{equation}
d^4=\frac{2^6}{3\pi} \frac{\rho g}{E\theta} w\,t\,l^3\,.
\label{eq:int2}
\end{equation} 
Injecting Eq. (\ref{eq:int1}) in Eq. (\ref{eq:int2}) allow to eliminate $t$, 
\begin{equation}
d^2=\frac{2^{13/4}}{(3\pi)^{1/2}} \Theta^{-3/4} w^{5/4}\,l^{3/2}\,.
\label{eq:int3}
\end{equation} 

Using those two intermediary equations, Eq.(\ref{eq:int1}) and (\ref{eq:int3}), in the equation of volume, Eq. (\ref{eq:vol}), the volume writes
\begin{equation}
 V=2^{3/2}\Theta^{-1/2} w^{5/2}\, l+\frac{2^{5/4}\pi^{1/2}}{3^{1/2}}\,\Theta^{-3/4} w^{5/4}\,l^{5/2}\,.
\label{eq:int_vol}
\end{equation} 

Now to find the optimal leaf, we have to maximize the leaf surface $S=2\,l\,w$ for a given volume $V$. We use the Lagrange multipliers technique where $\lambda$ is a dummy variable 
\begin{eqnarray}
\nonumber S&=&2Lw-\lambda \left( 2^{3/2}\Theta^{-1/2} w^{5/2}\, l\right. \\
&&\qquad\quad +\left. \frac{2^{5/4}\pi^{1/2}}{3^{1/2}}\,\Theta^{-3/4} w^{5/4}\,l^{5/2}\right)\,,
\label{eq:int_maxS}
\end{eqnarray} 
that gives respectively, ${\partial S}/{\partial L}=0$ and ${\partial S}/{\partial w}=0$
\begin{eqnarray}
\nonumber 2 w =\lambda\left(\frac{2^{3/2}}{\sqrt{\Theta}} w^{5/2} + \frac{5\,2^{1/4}\pi^{1/2}}{3^{1/2}} \Theta^{-3/4} l^{3/2}w^{5/4}\right) \\
\nonumber 2 l = \lambda\left(5\sqrt{\frac{2}{\Theta}} w^{3/2}\,l +  \frac{5\,\pi^{1/2}}{2^{3/4} 3^{1/2}} \Theta^{-3/4} l^{5/2}w^{1/4}\right) 
\label{eq:int_lagrange}
\end{eqnarray}

With these two lines, we eliminate $\lambda$ and find the relationship between $l$ and $w$
\begin{equation}
w^{5/4}=\left(\frac{2\,3^2}{11\,5\,\pi}\right)^{1/4}\Theta^{1/4}\,l^{3/2}\,.
\label{eq:int4}
\end{equation}

Now, simple calculations with Eq. (\ref{eq:int1}), (\ref{eq:int3}), (\ref{eq:int4}) and (\ref{eq:int_vol}) give

\begin{equation}
l= \left(\frac{2\,3^2\, \theta}{5\,11\,\pi}\right)^{1/4}\left(\frac{E}{\rho g}\right)^{1/4} V^{1/4}\,,
\label{eq:scale_l2}
\end{equation}

\begin{equation}
w= \left(\frac{5^5\, \pi\, \theta}{2^7\,3^{3}\, 11^{3}}\right)^{1/10}\,\left(\frac{E}{\rho g}\right)^{1/10} V^{3/10}\,, 
\label{eq:scale_w2}
\end{equation}

\begin{equation}
d= \left(\frac{2^{11}\,3\,5}{11^3 \,\pi^3 \,\theta}\right)^{1/8}\left(\frac{E}{\rho g}\right)^{-1/8} V^{3/8} \,,
\label{eq:scale_d2}
\end{equation}

and finally

\begin{equation}
t= \left(\frac{\pi^3\, 5^{15}}{3^{9} \,2^{11} \,11^9\, \theta^7}\right)^{1/20}\left(\frac{E}{\rho g}\right)^{-7/20} V^{9/20}\,. 
\label{eq:scale_t2}
\end{equation}

\end{document}